\documentstyle[twocolumn]{article}
\def\section#1{\bigskip\centerline{\bf #1}\medskip}

\begin{document}

\title{\rightline{\small IISc-CTS-5/01}\vspace*{-0.4truecm}
       \rightline{\small quant-ph/0105001}
{\Large\bf Why genetic information processing could have a quantum basis}}

\author{\normalsize APOORVA PATEL\\
        \normalsize\it
        CTS and SERC, Indian Institute of Science, Bangalore-560012, India\\
        \normalsize\it
        (Fax, +91-80-3600106; E-mail, adpatel@cts.iisc.ernet.in)}

\date{\small
\leftline{
Living organisms are not just random collections of organic molecules.
There is continuous information processing going on}
\leftline{
in the apparent bouncing around of molecules of life.\ \ Optimisation
criteria in this information processing can be searched}
\leftline{
for using the laws of physics.\ \ Quantum dynamics can explain why living
organisms have 4 nucleotide bases and 20 amino}
\leftline{
acids, as optimal solutions of the molecular assembly process.
Experiments should be able to tell whether evolution indeed}
\leftline{
took advantage of quantum dynamics or not.}
\medskip\leftline{{\small\bf Keywords.}\hfill\small
Amino acid; computation; database search; DNA; enzyme; information;
nucleotide base; protein; quantum}
\leftline{\small
coherence; quantum mechanics; superposition; unitary evolution}
\medskip\hrule}

\maketitle

\section{1. Information}

What is life? About fifty years ago, Erwin Schr\"odinger attempted to answer
this question on the basis of known laws of physics (Schr\"odinger 1944).
His insight has since then inspired many researchers to investigate the
molecular basis of a living organism. Chemical bonds explain how atoms bind
together to form various molecules. It is possible to take the common
elements $H$, $C$, $N$, $O$, stir them together with some heat and electric
sparks, and obtain molecules of life such as water, methane, ammonia, sugars,
amino acids, nucleotide bases, and so on. These molecules exist even in the
interstellar clouds. It is also not difficult to arrange these molecules in
an orderly manner as in a crystal, or jumble them up in a random ensemble
as in a gas. But living organisms are neither ordered crystals nor random
mixtures of their building blocks. The building blocks of a living organism
are linked together in a precise fashion to make functional parts. These
links between building blocks are often indirect and not physical; they
describe an order amongst the building blocks. Information is the abstract
mathematical concept that quantifies the notion of this order amongst the
building blocks, and it was this concept that was emphasised by Schr\"odinger
in his seminal work.

It is easiest to quantify information using the framework of communication.
When a message is conveyed by one person to another, the measure of the
information contained in the message is the increase in the knowledge of the
second person upon receiving the message from the first. The larger the
number of possibilities for a message, more is the amount of uncertainty
removed upon its receipt, and so more is the information contained in it.
The simplest message would be just a yes or no, distinguishing amongst only
two possibilities. Claude Shannon thus defined the information contained in
a message as its entropy; it directly measures the number of possibilities
for the message. A repetitious message wastes resources repeating what is
already conveyed before. So the information contained in a message is
increased by removing correlations amongst its parts; as messages become
more efficient, they appear more and more random.

Information thus lives in randomness, but it is not randomness. What
distinguishes it from randomness is the sense of purpose, i.e. the message
has a meaningful interpretation for the receiver. Of course, that requires
a common language which both the sender and the receiver understand. Often a
prior agreement fixes the language, but then the agreement itself would have
been made by an earlier message. Going back all the way, one can ultimately
connect the interpretation of a message to the physical properties of the
objects that carry the message. The most primitive messages have to be of
this type, and then higher level sophisticated structures can be constructed
using them. Trying to figure out what the best language would be in a given
situation is too vast an area of investigation; here I concentrate only on
information in messages with a fixed language.

It is a characteristic of living organisms to acquire information,
interpret it and pass it on, often using it and refining it along the way.
This information can be in various forms. It can be genetic information
passed on from the parent to the offspring, sensory information conveyed
by the sense organ to the brain, linguistic information communicated by
one being to another, or numerical data entered in a computer for later use.
Living organisms are thermodynamically open non-equilibrium systems. They
absorb free energy, and use it to create order within and throw disorder out.
The ultimate source of this free energy has to be an interaction that is not
in equilibrium. In our world this interaction is gravity; gravity is always
attractive and so cannot be in equilibrium---the lowest energy state is a
black hole (see for instance, Davies 1998).

Computer science is the mathematical framework for processing information.
A computer takes certain information, in the form of an input, and by
suitable manipulations converts it into an output. The manipulations are
defined by mathematical algorithms and implemented by physical devices.
Obviously the types of manipulations that can be carried out are limited by
the types of physical devices available. Efficient computers are those
that reliably accomplish their tasks using the least amount of resources.
Considering the living organisms to be specialised supercomputers, we can
study how efficient they are in implementing their tasks.

\section{2. Optimisation}

Darwinian evolution, i.e. survival of the fittest, describes the adaptations
of living organisms to their environment. These adaptations have occurred
by trial and error explorations, and not as a direct optimal solution to
a mathematical problem. Nowadays we understand them in the language of
genetics. Genes contain the essential information (i.e. the programme) of
life; they tell the rest of the living cell what to do in what circumstances.
The circumstances are provided by the environment, while the genes determine
the responses. Faithful replication of the genes passes on the information
from one generation to the next. Once in a while, chance mutations alter the
responses of the genes. If the change is beneficial the organism improves its
chances of survival, and if the change is detrimental the organism fades away.
It is important to note that mutations are local fluctuations, and cannot
bring about large changes in one go. The new organism is always similar to
the old one, and not a completely different one. In principle, large changes
can be built up from local ones over a long time. On the other hand, small
changes cannot get one out of a local optimum. This latter feature is
responsible for the wide variety of life we observe, and even though
evolution has progressed over a long period, we cannot be sure that it has
always discovered the optimal adaptations. Globally optimal features can
be only those which are widespread in organisms living under different
conditions. One such instance is the language of the genetic information;
it has remained unchanged from ancient bacteria to modern human beings.
It is worth exploring to what extent it is an optimal adaptation to the
available resources and the physical laws, using the principles of computer
science. (I find this approach much more appealing than the frozen accident
hypothesis (Crick 1968).
The pioneering contribution along this lines is: von Neumann 1958.)

Optimisation of information processing is essentially driven by two
guidelines: (1) minimisation of physical resources (time as well as space),
and (2) minimisation of errors. These guidelines often impose conflicting
demands, but we have learnt how to tackle them in the process of building
powerful computers, and it is instructive to analyse that in some detail.

The first step in optimal representation of a message is to break it up in
small segments. This is called digitisation. Instead of handling a single
variable covering a large range, it is much easier to handle several
variables each spanning a smaller range. Information of the whole message
is maintained by putting together as many as necessary of the smaller
range variables, while the instruction set required to manipulate each
variable is substantially simplified. This simplification means that only
a limited number of processes have to be physically implemented, and only
a limited types of physical variables have to be handled, leading to high
speed computation\footnote{%
We learnt mathematical tables in primary school to carry out addition and
multiplication using the decimal system. In binary system used by our
computers, these tables are replaced by only two operations, XOR and AND.}.
Furthermore, choosing the smaller range variables to be discrete, and not
continuous, it is possible to correct small errors. Continuous variables
can drift, and it is not possible to figure out the extent of the drift.
Discrete variables based on continuous physical properties can drift too,
but they can be reset to the nearest discrete variable, eliminating the
error whenever the drift is small\footnote{%
For example, voltages and currents in electrical circuits are continuous
variables, but the transistors in digital computer circuits are used
only in their discrete saturated states. Small voltage fluctuations are
eliminated by resetting the transistors to their saturated states. Only
when the voltage fluctuation is large, the state of the transistor flips
and there is an error in the calculation.}.
These advantages of a simple instruction set and error correction are
so overwhelming that it has become customary to describe a message
containing information as an aperiodic chain of building blocks. Our
systems of writing numbers and sentences have such a structure. Genetic
information has also incorporated this optimisation step: DNA and RNA
chains use an alphabet of 4 nucleotide bases, while polypeptide chains
use an alphabet of 20 amino acids.

The second optimisation step is the packing of the information in a message.
Repetitive structures or correlations amongst different parts of a message
reduce the capacity of the message to convey information---part of the
variables are wasted in repeating what is already conveyed. Elimination of
correlations reduces the length of a message; the information content of a
fixed length message is maximised when all the correlations are eliminated
and each of the variables is made as random as possible. On the other hand,
processing errors in an efficiently packed message destroys information.
If correlations exist amongst different parts of a message, they can be
exploited to eliminate local disturbances and to reconstruct the correct
message. Both these features are used in our computers: files are compressed
without losing information for efficient storage, and parity checks are
routinely performed to detect and reconstruct spoilt data. How efficiently
the information should be packed, and how many correlations should be kept,
is a trade-off that depends on the error rate of the particular information
processing system\footnote{%
We use one form of language to communicate to adults, and another form to
talk to babies. On comparing them, we immediately notice that the ``baby
language'' is full of repetitive syllables---an insurance against high
communication loss.}.
Detailed analyses of DNA sequences have found little correlation amongst
the letters of its alphabet in the coding regions, although correlations
do exist in the non-coding regions
(see for example, Arneodo {\it et al.} 1995, Nandy 1996).
We have to marvel at the fact that evolution has achieved the close to
maximum entropy structure of coding the genetic information.

Selection of the number of letters in the alphabet is the third optimisation
step. It clearly depends on the task to be accomplished and the choices
available as building blocks. There must be at least two building blocks for
a linear chain to carry information; a periodic crystal of a single building
block carries little information. While choosing a large number of building
blocks reduces the overall length of the message, it also makes distinguishing
them from each other more difficult. The practical criterion for fast
error-free information processing is therefore to choose as many letters
in the alphabet as can be quickly distinguished from each other. Different
physical reasons are involved in the selection of building blocks of different
information processing systems: the decimal system arose from our learning
to count with our fingers, the number of syllables in our languages are
determined by the number of distinct sounds our vocal chords can make, binary
code is used in computers and nervous systems because off/on states can be
quickly decided with electrical signals. Are there any such physical
underpinnings for the number of letters chosen for DNA and protein chains?
The task involved in genetic information processing is ASSEMBLY. The desired
components already exist (they are floating around in a random ensemble);
they are picked up one by one and linked together in the required order.
Whether a particular component is the desired one or not is decided by
base-pairing, and it is a simple yes/no query---either the base-pairing
takes place or it does not.  The optimisation criterion for this task is now
clear: find the number of items that can be reliably distinguished from each
other given a fixed number of yes/no queries. This is a mathematically
well-defined problem, to be solved using the laws of physics available at
the molecular scale.

Many more than 4 nucleotide bases and many more than 20 amino acids can be
synthesised by chemical reactions. Indeed many of them exist in the cellular
environment. But DNA and RNA always contain 4 nucleotide bases and polypeptide
chains always contain 20 amino acids. A number of attempts have been made
to understand these mysterious numbers, since the structure of DNA was
unraveled. Since DNA replication required complementary base-pairing, it was
reasonable to expect the number of nucleotide bases in DNA to be even. Beyond
that there was no understanding of why the number of nucleotide bases should
be 4, except that 2 is the smallest possible value and 4 is the next one.
Far more elaborate schemes were constructed to explain why polypeptide chains
should contain 20 amino acids. They were based on stereochemical properties
of the molecules involved, various permutations of the nucleotide bases and
combinatorics (Hayes 1998 has a recent summary of these efforts).
All these schemes fell apart with the discovery of the non-overlapping triplet
code.

\section{3. Two Languages}

Before trying to figure out the optimal number of letters for the DNA and
protein alphabets, let us first understand why genetic information processing
requires two distinct languages---one with the nucleotide bases as the
building blocks and another with the amino acids as the building blocks.
We often translate one language into another by replacing one set of building
blocks with another. When the languages are versatile enough this translation
can be carried out without any loss in the information content; a code
specifies which set of building blocks are translated into which ones.
A particular set of building blocks is selected not by the abstract
information content that has to be conveyed, but by its suitability for
the physical tasks to be carried out during processing of the information.
Our computers compute using electrical signals but store the results on the
disk using magnetic signals; the former realisation is suitable for quick
processing while the latter is suitable for long term storage. Proteins and
DNA participate in similar tasks. Proteins are actively involved in many
biochemical processes going on in the cell, and suffer much wear and tear
as a result. The double helical structure of DNA, with the nucleotide bases
hidden inside, carefully protects the information until it is required.
Also, DNA replication is much less error-prone than protein synthesis.

This is only part of the story. Translation between DNA and protein languages
is more complicated than just swapping one set of building blocks for another.
When magnetic signals are converted into electric signals in a computer, the
building blocks change but the language does not; both forms represent the
same sequence of zeroes and ones. In the case of DNA and proteins, not only
the building blocks change, but the language undergoes a change too. The
translation would have been easier if only the building blocks changed as
per physical requirements, but not the language. Then what necessitated a
change in the language? The reason again has to do with the task associated
with the information. For example, the textual information typed on the
keyboard is stored in a computer in binary format using the ascii code. The
textual format is easier for humans to read and speak, while the binary
format is easier for computers to manipulate. The job assigned to DNA is
faithful replication and transcription, which is easily accomplished in the
form of one-dimensional chains of building blocks. The job assigned to
proteins is participation in biochemical reactions, where size and shape of
the protein play a critical role. The building blocks of proteins---the
amino acids---have to therefore know how to fold one-dimensional chains into
three-dimensional structures. The number of building blocks required to
encode one-dimensional chains and three-dimensional structures is certainly
different (Patel 2001b).
It is this drastic change in physical realisation of the information that
has driven the living organisms to develop two distinct languages, and the
complex machinery that translates one into the other.

There is yet another distinction between the languages of DNA and proteins.
The language of DNA is a high level one, analogous to the compact languages
we use to write our computer programmes. It is symbolic and abstract, and
it requires interpretation before the information contained in it can be
used. The language of proteins is a low level one, even more direct than
the machine codes which run computers. There is no other agency to interpret
the information; the building blocks themselves carry the instructions of
what is to be done in terms of their physical properties. The sequence of
amino acids ``knows'' how to fold itself, which in turn decides which
biochemical reaction the protein will participate in.

The concepts of information processing and optimisation have been in the
background of all attempts to understand the features of the genetic code,
but the actual code that was discovered did not show any obvious relations
to them. (Boolean logic with yes/no queries can only produce 2 and its
powers as the possible number of building blocks.) Since then biologists
have considered, by and large, the genetic code as a frozen accident of
history: it arose somehow and became such a vital part of life that any
change in it would be highly deleterious (Crick 1968).
The situation changed again, with the realisation that quantum logic can be
used for information processing as well, and its optimisation features are
different than those of classical Boolean logic. So we can go back and look
again at the same problem, incorporating all that we have learnt in the
framework of computer science over the years. The next section outlines
how quantum logic explains the mysterious numbers of the genetic code as
solutions to an optimisation problem (Patel 2000b; Patel 2000a).

\section{4. Quantum Logic}

Quantum mechanics describes our understanding of how atoms are built from
their constituents and how they interact with each other. Its dynamical
equations are defined in terms of amplitudes (or wavefunctions), and the
classical observation probabilities are obtained by squaring the amplitudes.
It is important to note that these dynamical equations are precise, the
probabilistic interpretation arises only when we convert the dynamical
amplitudes into classical observables. The amplitudes are a set of complex
numbers, normalised to the total probability being one, and they evolve in
time by unitary transformations. Complex amplitudes and their unitary evolution
follow a totally different dynamics compared to real probabilities and
their Markovian evolution. This difference lies at the basis of why quantum
algorithms and their optimisation criteria are distinct from their Boolean
counterparts. It also makes the quantum logic is superior to the classical
one---the same task is accomplished using less resources (time and space).

The ASSEMBLY process is a variation of the SEARCH process, where the desired
object is picked up from an ensemble based on certain property checks. In
case of genetic information processing, the ensemble is a random one and the
property checks are implemented by molecular bonds involved in base-pairing.
The dynamics of molecular bond formation is no doubt quantum, and so a
quantum search algorithm is a possibility for genetic information processing.

The optimal quantum search algorithm was found by Lov Grover (Grover 1996),
and it relates the number of objects, $N$, that can be distinguished by a
number of yes/no queries, $Q$, according to
\begin{equation}
(2Q+1) \sin^{-1} (1/\sqrt{N}) = \pi/2 ~.
\end{equation}
This algorithm does not use the full power of quantum logic; concepts of
superposition and interference familiar from the study of classical waves
are sufficient to describe it. The algorithm starts with a uniform 
superposition of all possible states, corresponding to equal probability
for every building block to get selected. Then it applies two reflection
operations alternately: (a) change the sign of the amplitude of the
desired state by the yes/no query, and (b) reflect all amplitudes about
their average value. The algorithm stops after $Q$ of these alternating
reflections to yield the desired state with a high probability.

The solutions of Eq.(1) for small values of $Q$ have special significance
for the number of building blocks involved in genetic information processing
(details of the genetic code can be found in Watson {\it et al.} 1987;
Lewin 2000):
\begin{eqnarray}
Q=1 \Longrightarrow& N=4    ~, \cr
Q=2 \Longrightarrow& N=10.5 ~, \cr
Q=3 \Longrightarrow& N=20.2 ~.
\end{eqnarray}
(1) A single base-pairing distinguishes between 4 possibilities in DNA
replication and m-RNA transcription. This is an exact solution of Eq.(1),
so chances of error are minimised.\\
(2) Bilingual t-RNA synthetases ensure the matching between the amino acid
at one corner of the t-RNA molecule and the anticodon triplet at another.
These synthetases belong to two distinct classes of 10 each, distinguished
by the structure of their active sites and by how they attach amino acids
to the t-RNA molecules. The lack of any relationship between the two classes
have led to proposals that they evolved independently. It is quite plausible
that early forms of life existed with proteins that were made up of just 10
amino acids, belonging to one class or the other and coded by two nucleotide
bases. The wobble rules and similar codons in the genetic code for amino
acids with similar properties would then be relics of the merger of the
two distinct classes during evolution (Patel 2001b). 
$N=10$ is not an exact solution for $Q=2$, which means that the quantum
algorithm will not always find the desired object. There exists a small
probability, about 1 part in 1000, that the quantum algorithm will select
a wrong object.\\
(3) Three base-pairings between t-RNA and m-RNA transfer the information
from the nucleotide base chain to the amino acid chain. The non-overlapping
triplet genetic code carries 21 signals (20 for the amino acids plus a STOP)
in this process. $N=21$ is not an exact solution for $Q=3$; the quantum
algorithm then has an intrinsic error probability of about 1 part in 1000.

\section{5. DNA Structure}

Having discovered that the optimal quantum search algorithm can explain
the number of building blocks involved in genetic information processing,
the next exercise is to look for the physical implementation of the steps of
the algorithm. During DNA replication, the intact strand of DNA acts as a
template on which the growing strand is assembled. At each step, the base
on the intact strand decides which one of the four possible bases in the
environment can pair with it. This is exactly the yes/no query used in the
search algorithm. Based on the known features of this process, I have
proposed the following scenario (Patel 2000a):\\
(a) The molecular bonds involved in base-pairing are Hydrogen bonds. They
can be explained only using the language of quantum mechanics, and correspond
to quantum tunnelling of a proton ($H^+$) between two attractive energy minima.
When the nucleotide bases come together with random orientations, their pairing
takes place in a two-step process. The formation of the first bond still
leaves enough freedom for nucleotide bases to rotate and orient in various
ways in the three-dimensional space. The second step fully locks the nucleotide
bases in their bound structure. Such a two-step base-pairing process has the
correct quantum dynamics to flip the sign of the amplitude of the desired
state.\\
(b) Over a long time, the quantum amplitudes relax towards an equilibrium
state. The base-pairing takes place on a very short time scale, and acts
as a sudden disturbance. The amplitudes then again try to relax back to the
equilibrium state, just like a damped pendulum which is suddenly kicked.
The opposite end of the oscillation of the amplitudes about the equilibrium
state corresponds to the reflection about average operation. If the quantum
algorithm is stopped there by extracting the binding energy, the desired
base-pairing is achieved.\\
(c) The job of creating the uniform superposition of all states, and then
maintaining the coherence of the quantum dynamics is assigned to the enzymes.
Enzymes play a crucial catalytic role in the replication process; the process
simply does not occur in the absence of enzymes by chance molecular collisions.
Disturbances from the environment, called decoherence, are extremely fast and
generally destroy quantum features of macroscopic systems in no time. Enzymes
do provide shielded environments at the molecular scale, and modern techniques
of molecular biology can be used to check whether they can offset the effects
of decoherence in the case of DNA replication or not (Patel 2001a).

\section{6. Future}

I have analysed the molecular assembly process from the view-point of
information theory and optimisation. The best algorithm for accomplishing
this task is based on quantum dynamics. The optimal number of building
blocks predicted by the algorithm agree remarkably well with the number of
building blocks involved in genetic information processing---replication
of DNA and synthesis of proteins. The molecular structure and dynamics of
base-pairing also has the features necessary to implement the quantum
algorithm. The question that remains is: does genetic information processing
really use the quantum algorithm, or is the existence of both software and
hardware features an accident?

This genetic information processing takes place at the atomic scale, where
quantum mechanics is the framework for understanding the physical processes.
It is reasonable to expect that if there was something to be gained from
quantum algorithms, life would have taken advantage of that at this physical
scale. Of course, the structure of DNA came into existence billions of years
ago, and it could be that what was relevant when life arose is not relevant
now; the observed features could be just left-overs from a bygone era. Even
though we cannot recreate the conditions in which life originated, we can
experimentally test whether at present quantum dynamics plays a role in
genetic information processing or not (Patel 2001a).

The biggest obstacle to quantum dynamics is decoherence, and it is worthwhile
to investigate whether living organisms have conquered it in at least some
of the fundamental processes of life or not. Experiments should be able to
check the quantum scenario for genetic information processing, and then we
would know the adequacy of our current understanding of molecular biology.

Comparing genetic information processing to our modern digital computers,
we can observe that DNA plays the role of memory, m-RNA plays the role
of registers, enzymes play the role of instructions, while ribosomes and
other complicated structures in the cell carry out the tasks of the CPU.
It is the CPU and the instructions that make it work, that form the heart
of the computer. Memory is the simplest, and quite likely the last to be
developed, component of the computer. Using the language of information
theory, therefore, I have looked at only the simplest of the processes.
There are deeper questions to be addressed, and more complicated processes
to be analysed. For example, what kind of optimisation selected particular
physical objects as the building blocks? What kind of primitive machinery
could have led to the present genetic code? What is the relevance of the
degeneracy of the genetic code for amino acids? What is the dynamics behind
the catalytic role played by the enzymes? All that is for the future.
I can only say that the possible involvement of quantum dynamics in genetic
information processing has provided a novel way of analysing these processes.

\section{Acknowledgements}

\noindent
I am grateful to S. Mahadevan, V. Nanjundiah and H. Sharatchandra for their
helpful suggestions.

\section{Appendix}

Here I describe the meanings of some of the technical words used in this
article.\\
{\bf Superposition:} For linear dynamical systems, a linear combination of
possible solutions is also a solution to the same dynamics. For convenience,
a suitable basis is chosen, and a general solution is parametrised by its
components along the basis directions. In wave mechanics and non-relativistic
quantum dynamics, these components are complex numbers and are called
amplitudes. Superposition corresponds to addition of the amplitudes when
combining several solutions. Complex amplitudes can add constructively as
well as destructively, leading to interference patterns. Observation
probability of a particular solution is given by the absolute square of
the corresponding amplitude.\\
{\bf Unitary evolution:} A linear dynamical system evolves by a linear
transformation of its components, often denoted as multiplication of a
vector by a matrix. Conservation of total observation probability implies
that the squared-norm of the vector is always unity, and the most general
transformation is a rotation. When the components of the vector are complex
numbers, the most general evolution matrix is a unitary one.\\
{\bf Markovian evolution:} Classical probability theory can also describe
linear dynamics, with individual probabilities assigned to every component.
Each component probability has to be a real number between zero and one,
and the sum of all components is always unity. The elements of the evolution
matrix are real numbers between zero and one, such that elements in every
column of the matrix add up to one. Such an evolution of probabilities is
called Markovian evolution.\\
{\bf Decoherence:} Although quantum dynamics provides a perfect description
of processes occurring at the atomic scale, we hardly observe any quantum
effects at the macroscopic scale. Decoherence is the explanation of how
interaction of a quantum system with its environment can reduce its quantum
behaviour to classical one. Collisions and scatterings carry away complex
phases of the quantum system into the surrounding environment, where they
are irretrievably lost. A quantum system without its complex phases cannot
be described in terms of superposition of amplitudes; it has to be described
in terms of ``averaged'' statistical probabilities. Decoherence can be
reduced by insulating the quantum system from its environment, but that
becomes highly difficult as the system size increases. In general, decoherence
is extremely fast even for microscopic systems---so fast that it has not been
directly observed.

\bigskip
\centerline{\bf References}
\medskip

{\small
\noindent
Arneodo A, Bacry E, Graves P V and Muzy J F 1995
Characterizing long range correlations in DNA sequences from wavelet analysis;
{\it Phys. Rev. Lett.} {\bf 74} 3293-3296

\noindent
Crick F H C 1968 The origin of the genetic code;
{\it J. Mol. Biol.} {\bf 38} 367-379

\noindent
Davies P 1998 {\it The fifth miracle} (London: Penguin books)

\noindent
Grover L 1996 A fast quantum mechanical algorithm for database search;
{\it Proc. 28th Annual ACM Symposium on Theory of Computing} (Philadelphia)
pp 212-219 [quant-ph/9605043]

\noindent
Hayes B 1998 The invention of the genetic code;
{\it American Scientist} {\bf 86} 8-14

\noindent
Lewin B 2000 {\it Genes VII} (Oxford: Oxford university press)

\noindent
Nandy A 1996 Graphical analysis of DNA sequence structure:
III. Indications of evolutionary distinctions and characteristics of
introns and exons; {\it Curr. Sci.} {\bf 70} 661-668

\noindent
Patel A 2000a Quantum algorithms and the genetic code;
{\it Proc. Winter Institute on Foundations of Quantum Theory and Quantum
Optics} (Calcutta), {\it Pram{\=a}{\d n}a} {\bf 56} 367-381
[quant-ph/0002037]

\noindent
Patel A 2000b Quantum database search can do without sorting;
{\it Preprint} IISc-CTS-1/00 [quant-ph/0012149]

\noindent
Patel A 2001a Testing quantum dynamics in genetic information processing;
{\it J. Genet.} {\bf 80} (in press) [quant-ph/0102034]

\noindent
Patel A 2001b Carbon---the first frontier of information processing;
{\it Preprint} IISc-CTS-8/01 [quant-ph/0103017]

\noindent
Schr\"odinger E 1944 {\it What is life?}
(Cambridge: Cambridge university press)

\noindent
von Neumann J 1958 {\it The computer and the brain}
(New Haven: Yale university press)

\noindent
Watson J D, Hopkins N H, Roberts J W, Steitz J A and
Weiner A M 1987 {\it Molecular biology of the gene}, 4th edition
(Menlo Park: Benjamin/Cummings)
}

\end{document}